\documentclass{SCGE}
\usepackage{mathrsfs}
\usepackage{graphicx}
\begin{document}

\begin{picture}(0,0){\rm
\put(0,-39){\makebox[160truemm][l]{\bf {\sanhao\raisebox{2pt}{.}}
Article  {\sanhao\raisebox{1.5pt}{.}}}}}
\put(0,-52){\jiuwuhao {\textcolor[rgb]{0.5,0.5,0.5}{\sf 
}}}
\end{picture}
\def\bm{\boldsymbol}

\def\dl{\displaystyle}
\def\du{\end{document}}
\def\pi{{\uppi}}
\def\km{\,{\rm km}}
\def\cm{\,\rm cm}
\def\s{\,{\rm s}}
\def\e{\,{\rm e}}
\def\mpc{\,{\rm Mpc}}
\def\erg{\,{\rm erg}}
\def\kev{\,{\rm keV}}
\def\mev{\,{\rm MeV}}
\def\gev{\,{\rm GeV}}
\def\ergs{\,{\rm erg}}
\def\mum{\,{\rm \mu\rm m}}
\def\yr{\,{\rm  yr}}
\def\Gpc{\,{\rm Gpc}}
\def\msun{\,{M_\odot}}

\newcommand{\mes}{M\'esz\'aros}

\Year{2013} %
\Month{May} %
\Vol{56} 
\No{5} 
\BeginPage{1029} 
\EndPage{1034} 
\AuthorMark{{{\rm Tan W W.} et al.}}  
\AuthorMarkCite{{ }.} 
\DOI{10.1007/s11433-013-5079-4} 


\title{Implications of the cosmic infrared background excess for the cosmic star formation}
\author{Tan Wei-Wei \& Yu Yun-Wei$^*$}{}

\address[{\rm}]{Institute of Astrophysics, Central China Normal University,
Wuhan, China}
\maketitle \vspace{-3.5mm}{\footnotesize\begin{center} Received March 25, 2012; accepted May 25, 2012; published online April 15, 2013
\end{center}}\vspace*{-5mm}

\begin{center}
\rule{16.5cm}{0.4pt}
\parbox{16.5cm}
{\begin{abstract}   By phenomenologically describing the
high-redshift star formation history, i.e.,
$\dot{\rho}_{*}(z)\propto[(1+z)/4.5]^{-\alpha}$, and
semi-analytically calculating the fractions of high-redshift Pop
I/II and Pop III stars, we investigate the contributions from both
high-redshfit Pop I/II and Pop III stars to the observed
near-infrared ($3~ \mu\rm m<\lambda<5~\mu m$) excess in the cosmic
infrared background emission. In order to account for the
observational level of the near-infrared excess, the power-law index
$\alpha$ of the assumed star formation history is constrained to
within the range of $0\lesssim\alpha\lesssim1$. Such a constraint is
obtained under the condition that the virial temperature of dark
matter halos belongs to the range of $500{~\rm K}\leq T_{\rm
vir}\leq10^4$ K.
\end{abstract}}

\end{center}\vspace*{-0.6cm}

\begin{center}
\parbox{16.5cm}
{\bf\jiuhao infrared background, star formation, dark matter}
\end{center}

\begin{center}
{}
\Cit{Tan W W, Yu Y W. Implications of the cosmic infrared background
excess for the cosmic star formation. Sci China-Phys Mech Astron,
2013, 56: 1029-1034, doi: 10.1007/s11433-013-5079-4
}
\end{center}

\wuhao\vspace*{1.5mm}

\begin{multicols}{2}

\renewcommand{\baselinestretch}{1.08} \baselineskip 12.2pt\parindent=10.8pt

\renewcommand{\thefootnote}

\section{Introduction}
\no In recent years, the cosmic star formation history (CSFH) is of
more and more interest in both astrophysical and cosmological
studies, especially the early universe where direct detections are
difficult and scarce. As representative objects in the early
universe, the first generation stars, commonly called Population III
(hereafter Pop III) stars, would play a key role in the early
evolution of the cosmic structure. Pop III stars are thought to
originate from primordial metal-free gas at redshifts exceeding
$z\sim 10$ and precede the normal metal-enriched stellar populations
(see References \cite{Bromm04,Glover05, Bromm09} for reviews of Pop
III stars). Direct observations of Pop III stars with current
telescopes are impossible due to their high redshifts, although
these stars are very massive and luminous. Alternatively, Pop III
stars are expected to have left a great amount of diffuse radiation
in the universe, which is shifted today into the infrared
wavelength. For instance, for a photon at Lyman-$\alpha$ limit with
$\lambda_s=912~\rm \AA$ emitted from $z=20$, its wavelength to be
observed will be redshifted to
$\lambda_{r}=\lambda_s(1+z)=1.9~\mu\rm m$. Therefore, a significant
near-infrared

\vspace*{-3mm}
\noindent\rule{2.5cm}{0.4pt}\\[0.1mm]{\qihao *Corresponding author (email:
yuyw@phy.ccnu.edu.cn)}

\no component from Pop III stars is expected to be present in the
cosmic infrared background (CIB).

In the aspect of observation, a substantial near-infrared excess
over the net fluxes produced by galaxies has been indeed inferred
from the measurements of CIB anisotropies with COBE/DIRBE
\cite{Hauser98}, Deep Spitzer Infrared Array Camera (IRAC)
\cite{Kashilinsky04, Kashilinsky07}, Infrared Telescope in Space
(IRTS) \cite{Matsumoto05}, and Hubble Ultra Deep Field (HUDF)
\cite{Thompson07}. The commonly known bolometric flux of such near
infrared background excess (NIRBE) reads $F_{\rm excess}^{\rm
obs}=(2.9\pm1.3)\times10^{-5}\rm~erg~s^{-1} cm^{-2} sr^{-1}$
\cite{Kashilinsky05}. However, a more recent analysis by
\cite{Thompson07, Arendt10} claimed that the flux of the NIRBE
within the wavelength range of $3~\mu\rm m<\lambda<5~\mu m$ could be
ten times smaller as
\begin{equation}
F_{\rm excess}^{\rm obs}\approx1\times10^{-6}\rm~erg~s^{-1} cm^{-2}
sr^{-1},\label{NCIB_EXCESS}
\end{equation}
because much of the previously estimated excess could be due to
inaccurate zodiacal light modeling \cite{Dwek05}.

Therefore, in this paper, we would use the observed NIRBE to
constrain the formation history of Pop III stars as well as the
properties of the first collapsing dark matter halos hosting such
stars. These halos may be too faint to be observed with the current
telescopes. Different from some previous research studies, in our
considerations, we also take into account the contribution to the
NIRBE from normal Pop I/II stars at high redshifts, because in the
process of derivation of the NIRBE flux, only the emission from the
galaxies at relatively low redshifts $z\lesssim 10$ was subtracted
\cite{Thompson07}. In the next section, we would present our
theoretical consideration of the formation rates of Pop III and Pop
I/II stars, where several free parameters are introduced. In section
III, by confronting the theoretical NIRBE with the observational
NIRBE, we constrain the model parameters. Finally, a summary and
discussion are given in the last section.

\section{Star formation rates}
In principle, the formation rate of Pop III stars can be
theoretically derived in the hierarchical formation model
 (e.g., \cite{Haiman99,Haiman06}), which however involves many complicated astrophysical
issues. Therefore, in order to decrease the model uncertainties and
utilize some observational information, we here give a
phenomenological description for the CSFH and a semi-analytical
calculation for the fraction of Pop III stars.

\subsection{Cosmic star formation history}
Following a series of measurements of star formation rate density,
especially the complication of Hopkins \& Beacom \cite{Hopkings06},
a consensus on the CSFH now emerges up to redshift $z\sim3.5$, which
includes a steady increase of star formation from $z=0$ to $z=1$,
and a following plateau up to $z\sim3.5$. An empirical fitting can
be written as
\begin{equation}
\dot{\rho}_*(z)\propto\\
\left\{
\begin{array}{lcl}\left({1+z}\right)^{3.44}, & {\rm
for}~
z\leq0.97,\,\\
\left({1+z}\right)^{0}, & {\rm for~} 0.97 < z\lesssim 3.5,\,
\end{array}\right.
\label{SFR_TOTAL}
\end{equation}
with a local rate density $\dot{\rho}_*(0)=0.02~ M_\odot ~\rm
yr^{-1} Mpc^{-3}$. In contrast, for higher redshifts, the situation
is still ambiguous. It could continue to plateau, drop off, or even
increase, e.g., as considered in \cite{Daigne06}. Therefore, we here
simply introduce a new free parameter $\alpha$ to parameterize the
high-redshift CSFH as follows
\begin{equation}
\dot{\rho}_*(z)\propto\left({1+z}\right)^{-\alpha},  ~{\rm for~}
z\gtrsim 3.5.\label{SFHhz}
\end{equation}
The power-law assumption adopted above is motivated by the shapes of
the history at $z\lesssim3.5$ (see equation \ref{SFR_TOTAL}) and
also implied by some preliminary measurements of the high-redshift
star formation rates through Lyman break galaxies \cite{Bouwens2008}
and through gamma-ray bursts as well as their host galaxies \cite{
Yuksel2008, Kistler2009, Ishida11, Elliott2012}.

As a phenomenological consideration, we further assume that the star
formation rate density given by equation (\ref{SFHhz}) has contained
both the contributions from Pop I/II and Pop III stars.

\subsection{Fraction of Pop III stars}
In the hierarchical formation model, star formation takes place
during the collapse and merging of dark matter halos. The formation
of the halos hosting zero-metallicity Pop III stars should satisfy
two fundamental conditions: (i) The halos must be ``freshly" formed
through collapse of diffuse dark matter but not through merging of
smaller halos. In these new halos, the stars are formed for the
first time. (ii) The newly formed halos are not located in the wind
radius, $R_{w}$, of old galaxies, because the intergalactic medium
within $R_{w}$ has been metal enriched by the galactic winds (i.e.,
feedback effect) \cite{Aguirre01, Madau01, Furlanetto03}. Here the
wind radius $R_{w}$ is defined as the Sedov radius of the galactic
wind, which is mainly determined by the total energy of all
supernovae in the galaxy. For typical energy release per supernova
$E_{SN}\sim10^{52}\rm erg$ and typical mass of Pop III stars
$M\sim100M_{\odot}$, we can have $R_{w}\sim
380~(m/10^{10}M_{\odot})^{1/5}[(1+z)/10]^{-3/5}{~\rm kpc}$.

For the first condition, following Furlanetto \& Loeb
\cite{Furlanetto05}, the fraction of matter that can collapse into
``new" halos can be expressed by
\begin{equation}
f_{\rm
new}(z)\approx\int_{\infty}^{z}\frac{m^2_{\min}}{\bar{\rho}}{\delta_c(z')\over\sigma(m_{\min})}n(m_{\min})\left|
\frac{d}{dz'}{\delta_c(z')\over\sigma(m_{\min})}\right|dz',
\end{equation}
where $\bar{\rho}$ is the mean cosmic density, $\delta_{\rm c}(z)$
is the critical density for collapse, $\sigma(m)^2$ is the variance
of the density field on the mass scale $m$, $n(m)$ is the Press \&
Schechter mass function \cite{PS}, and $m_{\min}$ is the minimum
halo mass. With a given virial temperature $T_{\rm vir}$, the
minimum halo mass can be estimated by equating the gravitational
force of the halos to the opposing gas pressure \cite{Barkana01}
\begin{equation}
m_{\rm min}(z)\approx10^8 M_\odot \left({\mu \over
0.6}\right)^{-3/2}\left({T_{\rm vir}\over 10^4~\rm
K}\right)^{3/2}\left({1+z\over10}\right)^{-3/2},
\end{equation}
where $\mu$ is the mean molecular weight, with $\mu=0.6$ for ionized
gas and $\mu=1.22$ for neutral primordial gas. Here we choose
$\mu=1.22$ as the universe is mostly neutral at high redshifts.

For the second condition, the probability that a new halo lies
within $R_w$ of an existing galaxy can be approximated by
\cite{Furlanetto05}
\begin{equation}
p_e\approx1-\exp\left[-(1+B)\int^{\infty}_{m_{\min}}\frac{m}{\bar{\rho}}\eta(m)n(m)dm\right],\label{Qe}
\end{equation}
where $\eta(m)$ is the ratio of the mass metal-enriched by galactic
winds to the total mass of each galaxy. Following Furlanetto \& Loeb
\cite{Furlanetto05} , we have
\begin{equation}
\eta(m)\approx27 K_w
\left(\frac{m}{10^{10}M_{\odot}}\right)^{-2/5}\left(\frac{10}{1+z}\right)^{3/5},\label{wind}
\end{equation}
where the normalization factor $0\leq K_w\leq1/8$ accounts for many
feedback factors (e.g., radiative losses from supernova shocks). A
higher value of $K_w$ represents stronger feedback effects. For the
formation of Pop III stars, such feedback effect is a negative
effect. The coefficient $B$ appearing in Equation (\ref{Qe}) gives
the excess probability that two galaxies sit near each other, by
considering that the collapsing halos are actually biased due to the
existing halos and inclined to be near the existing halos. To be
specific, $B\approx \zeta_{\rm mm}(\bar{R}_w)b(m_{\min})\langle
b(m)\rangle$, where $\zeta_{mm}(\bar{R}_{w})$ (with $\bar{R}_w$ to
be

\begin{figure}[H]
\centering \resizebox{\hsize}{!}{\includegraphics{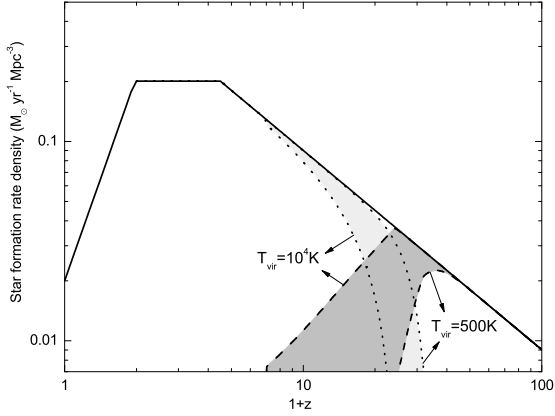}}
\caption{The CSFHs of Pop III stars (dashed lines), Pop I/II stars
(dotted lines), and their sum (solid lines), where $\alpha=1.0$ and
$K_w=0.02$ are taken. The shaded region represents the variation of
the virial temperature from $T_{\rm vir}=500~\rm K$ to $10^4~\rm
K$.} \label{fig1}
\end{figure}

\no the average wind size in the comoving units) is the dark matter
correlation function \cite{Eisenstein98}, $b(m)$ is the bias
function \cite{Mo02}, and the bracket $\langle \cdots\rangle$
represents an average on the Press \& Schechter mass function
\cite{Holzbauer12}.

\subsection{Formation rates of Pop III and I/II stars}
Combining the two aspects addressed above, the formation rate
density of Pop III stars as a function of $z$ can be given by
\begin{equation}
\dot{\rho}_{\rm III}(z)=\dot{\rho}_{*}(z) \left({df_{\rm
new}/dz\over df_{\rm coll}/dz}\right)(1-p_e),
\end{equation}
where $f_{\rm coll}(z)$ is the fraction of matter that can collapse
into ``all" of halos. For the Press \& Schechter mass function, it
reads \cite{Furlanetto05}
\begin{equation}
f_{\rm coll}(z)={\rm erfc}\left[\frac{\delta_{\rm
c}(z)}{\sqrt{2}\sigma(m_{\min})}\right].
\end{equation}
Meanwhile, the formation rate density of Pop I/II stars can be
obtained by subtracting the rate density of Pop III stars from the
total star formation rate density
\begin{equation}
\dot{\rho}_{\rm I/II}(z)=\dot{\rho}_{*}(z)-\dot{\rho}_{\rm III}(z).
\end{equation}

In the model described above, three free parameters remain:
$\alpha$, $T_{\rm vir}$, and $K_w$. As theoretically considered, the
value of $T_{\rm vir}$ is in principle determined by the cooling
style of the halos. The radiative cooling through molecular hydrogen
would lead to $T_{\rm vir}\sim500 $~K \cite{Abel95, Temark97}, while
the atomic cooling gives $T_{\rm vir}\sim 10^4 $~K (e.g.,
 \cite{Haiman00, Ciardi00}). Hence the specific value of $T_{\rm vir}$ relies
on the fraction of molecular hydrogen. For the temperature range of
$500{~\rm K}\leq T_{\rm vir}\leq 10^4 $~K, we present a numerical
result of the CSFH in Figure 1, which indicates that: (i) the
formation of Pop III stars would dominate the cosmic star formation
for $z>(20\sim40)$, and (ii) higher virial temperature $T_{\rm vir}$
could extend the formation of Pop III stars to lower redshifts.

\section{NIRBE emission}
Instead of an elaborate fitting to observations, here we would use
the observed bolometric flux of the NIRBE (Equation 1) to constrain
the model parameters. In the following calculations, the
contributions from Pop III stars and high-redshift Pop I/II stars
are both taken into account.
\subsection{Pop III stars}
As massive, zero-metallicity stars, Pop III stars would radiate
close to the Eddington limit $L_{\rm edd}={(4\pi
Gm_pc/\sigma_T)}M\simeq1.3\times10^{38}(M/M_\odot)~\ergs~\s^{-1}$,
where $M$ is the stellar mass. Then the lifetime of Pop III stars
can be estimated by $t_{L}\simeq\varepsilon M c^2 /L_{\rm edd}\simeq
3.1\times 10^6$ yr, which is independent of the stellar masses,
where $\varepsilon=0.007$ is taken as a fiducial value. Here Pop III
stars are not considered for significant mass loss during their
lifetimes because of their zero metallicity. The black body
effective temperature of Pop III stars can be estimated to $T_{\rm
III}\sim 10^5~\rm K$ with the Eddington luminosity and the
theoretical Hertzsprung-Russel diagram of Pop III stars
\cite{Schaerer02,Tumlinson03}.

At redshift $z$, the density of the energy radiated from Pop III
stars can be expressed by
\begin{equation}
\mathscr{E}_{\rm III}(z)\approx{\varepsilon \dot\rho_{\rm III}(z)
c^2t_H(z)},\label{unitenergy}
\end{equation}
which is released within the duration of $t_L$, where $t_H(z)\approx
(2/3)H(z)^{-1}$ is the Hubble timescale at that redshift. In the
standard $\Lambda$-cold dark matter cosmology, the Hubble expansion
rate reads $ H(z)=H_0 \sqrt{(1+z)^3\Omega_m+\Omega_\Lambda}$.
Hereafter $H_0=71~\rm km~s^{-1}Mpc^{-1}$, $\Omega_{\Lambda}=0.73$,
and $\Omega_m=0.27$ are taken. Therefore, the bolometric flux of the
NIRBE contributed by Pop III stars can be calculated by
\begin{eqnarray}
F_{\rm excess}^{\rm III}={1\over4\pi}\int \mathscr{K}_{\rm
III}(z){\mathscr{E}_{\rm III}(z)/t_L\over 4\pi d_l(z)^2}dV_c(z),
\end{eqnarray}
where $\mathscr{K}_{\rm III}(z)$ represents the fraction of energy
that enters the near-infrared wavelength range of $3~\mu\rm
m<\lambda<5~\mu\rm m$, $d_l(z)$ and $dV_c(z)$ are luminosity
distance and comoving volume element, respectively. To be specific,
$ d_l(z)=c(1+z)\int_0^zH^{-1}(z')dz'$ and $ dV_c(z)=4 \pi d_c(z)^2 c
H^{-1}(z)dz$ with $d_c(z)=d_l(z)/(1+z)$. By considering that
$\Omega_{\Lambda}\ll \Omega_{m}(1+z)^3$ for high redshifts, we can
simplify the above equation to
\begin{eqnarray}
F_{\rm excess}^{\rm III}\approx{c~l_{\rm III}\over6\pi
H_0^2\Omega_m}\int {\mathscr{K}_{\rm III}(z)\dot\rho_{\rm III}(z)
\over (1+z)^5 }dz,\label{fiii}
\end{eqnarray}
where Equation (\ref{unitenergy}) is used and $l_{\rm III}=L_{\rm
Edd}/ M=6.3\times10^{4}~\rm erg~s^{-1}g^{-1}$ represents the
luminosity per unit stellar mass.

\begin{figure}[H]
\centering \resizebox{\hsize}{!}{\includegraphics{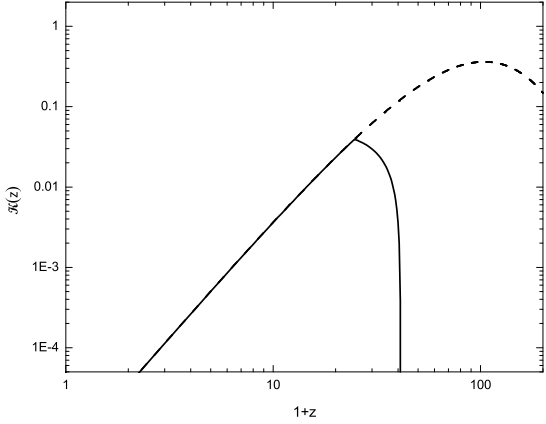}} \caption{
The dependence of the correction factor $\mathscr{K}(z)$ on redshift
for stellar temperature $T=10^5$ K (solid line), which peaks at
$z_p=24$ due to Lyman-$\alpha$ absorption. The case without
Lyman-$\alpha$ absorption is also presented by the dashed line for
comparison.} \label{fig2}
\end{figure}
For a black body spectrum described by the Planck function
$B_T(\lambda)$ for temperature $T$, the correction factor
$\mathscr{K}(z)$ can be generally calculated by
\begin{eqnarray}
\mathscr{K}(z)={\int_{\lambda_1/(1+z)}^{\lambda_2/(1+z)}B_T(\lambda)d\lambda\over\int_{0}^{\infty}B_T(\lambda)d\lambda}\label{kz1}
\end{eqnarray}
with $\lambda_1=3~\mu \rm m$ and $\lambda_2=5~\rm\mu m$
\cite{Arendt10}. However, as revealed by WMAP, the reionization of
the universe completes at most as early as redshift $\sim10.5$,
before which the universe is neutral \cite{komatsu2011}.
Consequently, for stars at high redshifts, photons emitted from them
with wavelengths shorter than the Ly-$\alpha$ wavelength (i.e.,
$\lambda_e\leq1216$ {\AA}) must be all absorbed by the diffuse
neutral hydrogens (as the Gunn-Peterson trough in the spectra of
high-redshift quasars). Therefore, the lower limit of the integral
in the numerator of Equation (\ref{kz1}) needs to be changed to
$\max[\lambda_1/(1+z),1216~{\rm {\AA}}]$. In the extreme, if
$\lambda_2/(1+z)<1216~{\rm {\AA}}$, then we have $\mathscr{K}(z)=0$.

For Pop III stars with $T=10^5$ K, a numerical result of
$\mathscr{K}_{\rm III}(z)$ as a function of redshift is presented in
Figure \ref{fig2}. As is shown, it firstly increases as $(1+z)^3$ up
to $z_p={3~{\rm \mu m}/ 1216~{\rm \AA}}-1=24$, and then plunges to
zero at $z_{\max}={5~{\rm \mu m}/ 1216~{\rm \AA}}-1=40$. The peak
value of $\mathscr{K}_{\rm III}(z)$ at $z_p$ can be analytically
calculated by
\begin{eqnarray}
\mathscr{K}_{\rm III}(z_p)&=&{2\pi hc^2\over\sigma
T^4\lambda_c^5}{(1+z_p)^4\Delta\lambda\over \exp[(1+z_p)hc/k_B
T\lambda_c]-1},
\end{eqnarray}
where $\lambda_c=4~\mu \rm m$ and $\Delta\lambda=2~\mu \rm m$.
Moreover, due to $(1+z_p)\ll 4k_BT\lambda_c/hc$, the peak value can
be approximately expressed by
\begin{eqnarray}
\mathscr{K}_{\rm III}(z_p)\approx{\pi ck_B\Delta\lambda\over\sigma
\lambda_c^4}{1\over T^{3}}(1+z_p)^3,\label{kiii}
\end{eqnarray}
which is proportional to $T^{-3}$ for $T>2\times10^4$ K.

\begin{figure}[H]
\centering\resizebox{1.1\hsize}{!}{\includegraphics{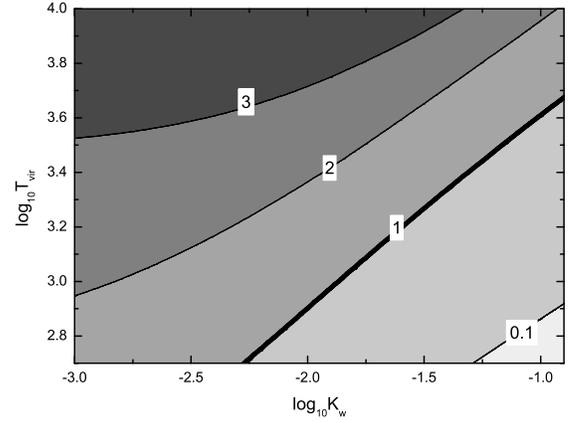}}
\caption{The distribution of the ratio $F_{\rm excess}^{\rm
III}/F_{\rm excess}^{\rm I/II}$ in the $K_w-T_{\rm vir}$ parameter
space.} \label{fig3}
\end{figure}

\subsection{Pop I/II stars}
Following Equation (\ref{fiii}), the contribution to the NIRBE
emission from high-redshift Pop I/II stars can be calculated by
\begin{eqnarray}
{dF_{\rm excess}^{\rm I/II}\over d\tilde{M}}={c~l_{\rm
I/II}\over6\pi H_0^2\Omega_m}\int{\mathscr{K}_{\rm I/II}(z)\over
(1+z)^5} {d\dot\rho_{\rm I/II}(z)\over d\tilde{M}}dz,\label{dfi/ii}
\end{eqnarray}
where $\tilde{M}\equiv M/M_{\odot}$ is the dimensionless stellar
mass. The lower limit of the above integral is set at $z\sim10$
since the contributions from Pop I/II stars at relatively low
redshift $z\lesssim 10$ have been subtracted from the excess
emission \cite{Thompson07}. The differential form is taken in
Equation (\ref{dfi/ii}) due to the wide distribution of the mass of
Pop I/II stars. So, different from Pop III stars, both the
quantities $l_{\rm I/II}$ and $\mathscr{K}_{\rm I/II}(z)$ here
should be treated as functions of stellar mass. Specifically, the
empirical mass-luminosity relation for Pop I/II stars tells us
$l_{\rm I/II}\approx\tilde{M}^{2.5}(L_{\odot}/M_{\odot})$.
Meanwhile, we can estimate the stellar temperature by
$T\sim5000\tilde{M}^{0.6}$ K according to the Hertzsprung-Russel
diagram of Pop I/II stars. For the Salpeter initial mass function,
$\phi(\tilde{M})\propto \tilde{M}^{-2.35}$, the differential star
formation rate can be expressed by
\begin{eqnarray}
{d\dot\rho_{\rm I/II}(z)\over d\tilde{M}}={A\dot\rho_{\rm
I/II}(z)\tilde{M}\phi(\tilde{M})},
\end{eqnarray}
where $A=\left[\int_{\tilde{M}_{\min}}^{\tilde{M}_{\max}}
\tilde{M}\phi(\tilde{M})d\tilde{M}\right]^{-1}=0.17$ with
$\tilde{M}_{\min}=0.1$ and $\tilde{M}_{\max}=100$. Then, we have
\begin{eqnarray}
F_{\rm excess}^{\rm I/II}&=&{Ac\over6\pi
H_0^2\Omega_m}\left({L_{\odot}\over M_{\odot}}\right)\nonumber\\
&&\times\int \tilde{M}^{1.15}\int{\mathscr{K}_{\rm
I/II}(z)\dot\rho_{\rm I/II}(z)\over
(1+z)^5}dzd\tilde{M}.\label{fi/ii}
\end{eqnarray}

In Figure 3, we plot the distribution of the ratio of $F_{\rm
excess}^{\rm III}$ to $F_{\rm excess}^{\rm I/II}$ in the $K_w-T_{\rm
vir}$ parameter space$^1$ \footnote{$^1$ The dependence of this
ratio on the parameter $\alpha$ is insensitive, because $\alpha$
influences the formation rates of Pop III and I/II stars in a same
way.}, which indicates that the contribution to the NIRBE from
high-redshift Pop I/II stars could be much more significant than
previously considered, especially for low $T_{\rm vir}$ and high
$K_w$.
\begin{figure*}
\resizebox{\hsize}{!}{\includegraphics{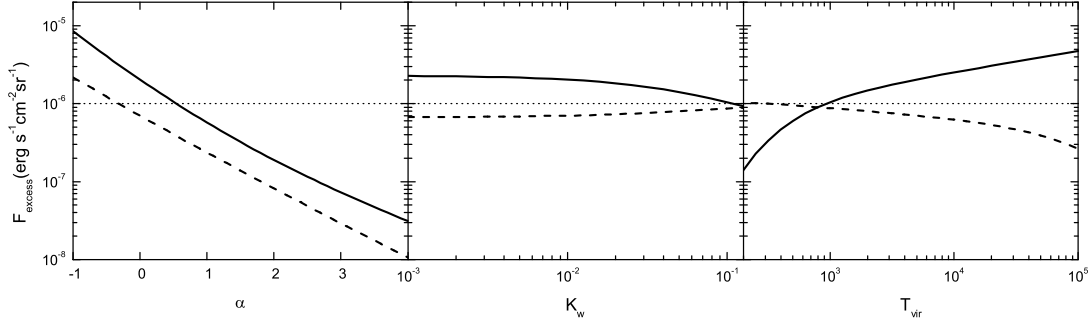}}
\caption{Dependence of $F_{\rm excess}$ on the free model parameters
for Pop III stars (solid lines) and high-redshift Pop I/II stars
(dashed lines), where the fiducial values of the parameters are
taken to $T_{\rm vir}=5000$~K, $K_w=0.01$, and $\alpha=0$. The
horizontal dotted line represents the observed flux of the NIRBE as
$1\times10^{-6}\rm~erg~s^{-1} cm^{-2} sr^{-1}$.}\label{fig4}
\end{figure*}

From Equation (\ref{fi/ii}), we can see that the value of $F_{\rm
excess}^{\rm I/II}$ could be mainly determined by the highest-mass
stars having temperatures $T>2\times10^4$ K. Therefore, following
Equation (\ref{kiii}), we can get
\begin{eqnarray}
{\mathscr{K}_{\rm I/II}(z)\over \mathscr{K}_{\rm
III}(z)}\approx\left({T_{\rm I/II}\over T_{\rm
III}}\right)^3\approx{\tilde{M}^{1.8}\over 8000},
\end{eqnarray}
where the peak redshift $z_p$ is replaced by an arbitrary redshift
$z$ because the profiles of the functions $\mathscr{K}(z)$ are
basically identical so long as $T>2\times10^4$ K. For a more direct
comparison with Pop III stars (Equation \ref{fiii}), we rewrite
Equation (\ref{fi/ii}) to
\begin{eqnarray}
F_{\rm excess}^{\rm I/II}&\approx&{c~{l'}_{\rm I/II}\over6\pi
H_0^2\Omega_m}\int{\mathscr{K}_{\rm III}(z)\dot\rho_{\rm
I/II}(z)\over (1+z)^5}dz,
\end{eqnarray}
where the equivalent luminosity per unit stellar mass reads
\begin{eqnarray}
{l'}_{\rm I/II}=A\left({L_{\odot}\over M_{\odot}}\right)\int
{\tilde{M}^{2.95}\over8000}d\tilde{M}=850~\rm erg~s^{-1}g^{-1},
\end{eqnarray}
which gives a ratio $ {l'}_{\rm I/II}/l_{\rm III}=0.014$. Therefore,
if $\dot\rho_{\rm I/II}$ is hundreds times higher than
$\dot\rho_{\rm III}$, the contribution to the NIRBE from
high-redshift Pop I/II stars can exceeds that from Pop III stars. As
exhibited in Figure 3, such a situation can arise with low $T_{\rm
vir}$ and high $K_w$.

\subsection{Constraining the CSFH}
The dependence of the model-predicted NIRBE fluxes on the parameters
$T_{\rm vir}$, $K_w$, and $\alpha$ are shown in Figure \ref{fig4}
for both Pop III and I/II stars. As can be seen, the uncertainty of
the NIRBE fluxes mainly comes from the uncertainties of $\alpha$ and
$T_{\rm vir}$, whereas the variation of $K_w$ only slightly
influences the value of $F_{\rm excess}$.

Combining Equations (\ref{fiii}) and (\ref{fi/ii}), the
model-predicted flux of the NIRBE emission contributed by both Pop
III and high-redshift Pop I/II stars can be calculated by
\begin{figure}[H]
\centering\resizebox{1.1\hsize}{!}{\includegraphics{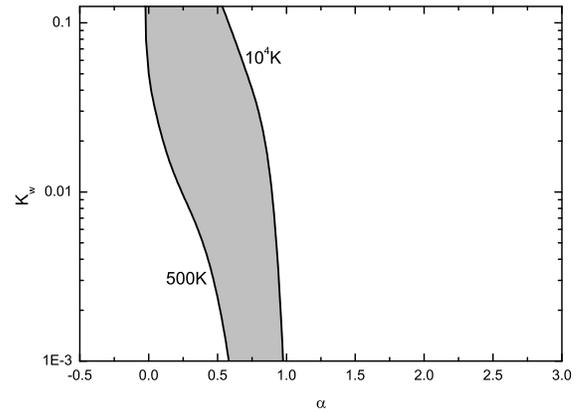}}
\caption{The parameter region where the model-predicted NIRBE flux
can account for the observational one (shaded region). The two
boundaries of the region corresponds to the virial temperatures of
500~K and $10^4$~K, respectively. } \label{fig5}
\end{figure}
\begin{eqnarray}
F_{\rm excess}^{\rm model}&=&F_{\rm excess}^{\rm III}+F_{\rm
excess}^{\rm I/II}.
\end{eqnarray}
\no As shown in Figure \ref{fig5}, by equating $F_{\rm excess}^{\rm
model}$ to $F_{\rm excess}^{\rm obs}$, we find a tight constraint on
the model parameters $\alpha$ as $0\lesssim \alpha\lesssim1$, where
the other two model parameters are considered to $500{~\rm K}\leq
T_{\rm vir}\leq 10^{4}$~K and $K_w\leq1/8$. This result indicates
that the high-redshift CSFH could be very flat. On the other hand,
specifically, lower virial temperature $T_{\rm vir}$ leads to lower
formation rates of Pop III stars, and thus the total formation rate
is required to be higher in order to have more Pop I/II stars to
produce emission.
\section{Summary and discussion}
In view of the inaccessible direct observations of high-redshift
stars, especially Pop III stars and their hosts, we propose an
indirect method to study these high-redshift objects by using the
observed CIB emission, particularly, the NIRBE emission. In our
model, the CSFH are described phenomenologically, in order to
decrease the complexity of the model. Consequently, a strong
constraint on the CSFH is found as $0\lesssim\alpha\lesssim1$, if
the virial temperature of the halos satisfies $500~{\rm K}\leq
T_{\rm vir}\leq 10^{4}$~K. Such a flat CSFH basically agrees with
the star formation rates inferred from high-redshift gamma-ray
bursts up to redshift $z\sim8-9$ \cite{Kistler2009}. In contrast,
for higher redshift (e.g., $z\sim10$), a recent measurement by
\cite{Bouwens11} claimed that the star formation rate could become
much lower. These observations imply that the realistic CSFH may
exhibit a behavior more complicated than a single power law. Anyway,
on the other hand, the CSFH with $0\lesssim\alpha\lesssim1$ predicts
sufficiently bright sources to reionize the universe to account for
the optical depth of cosmic microwave background photons
\cite{Yu12}, which is measured to be $\tau=0.088\pm0.015$ by WMAP.

The relationship between the parameters $\alpha$ and $T_{\rm vir}$
can in principle be derived from the hierarchal formation model,
although many uncertain astrophysical factors could be involved.
Combining such a theoretical relationship with the result presented
in this paper, we may obtain a more stringent constraint on the
value of $T_{\rm vir}$ as well as the CSFH.

\emph{We greatly acknowledge the anonymous referee for his/her
valuable comments, which have significantly improved our work. This
work is supported by the National Natural Science Foundation of
China (grant nos 11103004, 11073008, and 11178001) and by the
Self-Determined Research Funds of CCNU from the colleges’ basic
research and operation of MOE of China (Grant Nos. CCNU12A01010).}

\end{multicols}
\end{document}